\documentclass[a4paper,11pt]{article}
\usepackage{aaskaiid}
\usepackage{graphicx}
\usepackage{siunitx}
\usepackage{amsmath}
\usepackage{aas_macros}

\usepackage{aaskaiid,subfigure,orcidlink}

\title{Probing Merger Shocks in Galaxy Clusters in the SKA Era}
\ShortTitle{Cluster merger shocks}
\author[1,2]{Arpan Pal\orcidlink{0009-0007-8409-4233}}
\ShortName{Pal et al.} 
\author[1]{Ruta Kale\orcidlink{0000-0003-1449-3718}}
\author[3]{Gabriella Di Gennaro\orcidlink{0000-0002-8648-8507}}
\author[3]{Francesco de Gasperin\orcidlink{0000-0003-4439-2627}}
\author[4]{Swarna Chatterjee\orcidlink{0000-0001-8194-8714}}
\author[5]{Majidul Rahaman\orcidlink{0000-0002-1372-6017}}
\author[1]{Ramananda Santra\orcidlink{0009-0002-0373-570X}}
\author[6]{Mamta Pandey-Pommier\orcidlink{0000-0001-5829-1099}}
\author[7]{Abhirup Datta\orcidlink{0000-0002-5333-1095}}
\author[4,8]{Kenda Knowles\orcidlink{0000-0002-8452-0825}}
\author[7]{Nasmi S Anand\orcidlink{0009-0005-2553-6973}}
\affiliation[1]{National Centre for Radio Astrophysics, Tata Institute of Fundamental Research, S. P. Pune University Campus, Ganeshkhind, Pune 411007, India}
\affiliation[2]{National Radio Astronomy Observatory, 1011 Lopezville Road, Socorro, NM 87801-0387, USA}
\emailAdd{apal@ncra.tifr.res.in}
\emailAdd{ruta@ncra.tifr.res.in}
\emailAdd{ramananda1999@gmail.com}
\affiliation[3]{INAF - Istituto di Radioastronomia, via P. Gobetti 101, 40129 Bologna, Italy}
\emailAdd{g.digennaro@ira.inaf.it}
\affiliation[4]{Centre for Radio Astronomy Techniques and Technologies, Rhodes University, Drosty Rd, Makhanda 6140, South Africa}
\emailAdd{swarna.chatterjee@ru.ac.za}
\emailAdd{kenda.knowles@ru.ac.za}
\affiliation[5]{Institute of Astronomy, National Tsing Hua University, No. 101, Section 2, Kuang-Fu Road, Hsinchu 30013, Taiwan}\emailAdd{wrmajid@phys.nthu.edu.tw}
\affiliation[6]{Pole Scientific, University Catholic of Lyon, Campus Saint-Paul, 10 place des Archives 69288, Lyon Cedex 02, France}
\emailAdd{mamtapommier@gmail.com}
\affiliation[7]{Department of Astronomy, Astrophysics and Space Engineering, Indian Institute of Technology Indore, Indore, MP, India}
\emailAdd{abhirup.datta@iiti.ac.in}
\emailAdd{phd2301121010@iiti.ac.in}
\affiliation[8]{South African Radio Astronomy Observatory, Liesbeek House, Mowbray, Cape Town 7700, South Africa}

\abstract{Galaxy cluster mergers represent the most energetic phenomena in the Universe since the Big Bang releasing gravitational potential energy of $\sim 10^{63-64}$ erg, injecting turbulence and driving shocks through the intracluster medium (ICM). These merger shocks can accelerate cosmic ray electrons and compress magnetic fields, sometimes producing Mpc-scale synchrotron radio structures known as radio relics. 
Radio relics are powerful tracers of merger dynamics, particle acceleration, and magnetic field evolution, yet fundamental questions about the underlying physics and their time evolution remain unresolved. In this chapter, we review the current understanding of cluster merger shocks and their radio signatures, presenting the observational evidence from the SKA pathfinders and precursors, linking radio relics to shocks alongside outstanding theoretical challenges. 
We also consider related shock-influenced phenomena: radio phoenices from revived fossil AGN plasma and Gently Re-Energised Tails. 
Further, we outline the directions of investigation using the sensitivities of the SKA-Low and Mid complemented with X-ray observations that will allow us to make significant progress in understanding the cluster merger shocks. Detailed studies of individual targets in continuum and polarization and studies of populations of relics using wide surveys will both provide insights to the micro-physics and cosmic evolution of merger shocks.
We present SKAO capabilities across staged deployments starting from AA0.5 to AA4 and identify science verification targets that will illuminate the physics of cluster merger shocks in the SKA era.  
}

\begin{document}
\maketitle

\section{Galaxy cluster mergers}

In the hierarchical picture of structure formation, smaller structures assemble first and subsequently grow through successive mergers to build more massive systems \citep{1974ApJ...187..425P,1991ApJ...379...52W}. Galaxy clusters represent the largest gravitationally bound structures in the universe, comprising thousands of galaxies, diffuse gas, and dark matter with typical masses of $\sim 10^{14-15} \, M_{\odot}$. The mass budget of clusters is dominated by dark matter at approximately $85\%$ of the total mass, while diffuse gas contributes $10$--$12\%$ and galaxies comprise only a few percent. The diffuse gas component, termed the intra-cluster medium (ICM), is predominantly composed of thermal electrons and ions, cosmic rays, and magnetic fields \citep{1986RvMP...58....1S}. These electrons produce X-ray emission through thermal bremsstrahlung radiation. Over cosmic time, galaxy clusters grow via merging processes, which remain central to their ongoing evolution \citep{2005Natur.435..629S,2012ARA&A..50..353K}.

Galaxy cluster mergers rank among the most energetic phenomena in the Universe, releasing gravitational potential energy of order $10^{63-64}$ erg. During such mergers, the ICM undergoes violent perturbations characterized by turbulent flows and shock fronts \citep{2002ASSL..272....1S,markevitch07,2014ApJ...785..133H}. When subjected to these conditions, the plasma components undergo particle acceleration and magnetic field amplification, though the detailed physics remains poorly understood \citep[][]{2019SSRv..215...16V}.

Mergers are classified as major when colliding sub-clusters have comparable masses, and minor when a significantly smaller system merges with a much larger one. As one sub-cluster falls into another, density discontinuities develop in the ICM, forming shocks at the cluster periphery where particle densities are low. Binary mergers can produce axial shocks as well as equatorial shocks, though more complex multi-component mergers with varying viewing geometries present additional challenges for interpretation.

Due to the turbulent nature of the ICM, shock surfaces are inhomogeneous, composed of shocks with varying Mach numbers ($\mathcal{M}$)
rather than smooth discontinuities \citep[][Fig. \ref{fig:schematic}]{2009Ap&SS.322...65R, Ha2018}. The Mach number $\mathcal{M}$ is defined as the ratio between the shock velocity $v_{\rm shock}$ and the sound speed $c_s$ of the medium where the shock propagates.
As merger shocks expand outward from the core to the cluster outskirts, the average Mach number increases with time, reaching values of $\langle \mathcal{M} \rangle \sim 2$--$3$ \citep{2014ApJ...785..133H,Ha2018,2025ApJ...978..122L}. The kinetic energy flux through shocks peaks approximately 1 Gyr after their initial launching, at distances of $\sim 1$--$2$ Mpc from the core.

During hierarchical structure formation, even more powerful shocks with high Mach numbers emerge around and beyond the virial radii of clusters, termed accretion shocks \citep{1998astro.ph..5367E}. These are distinguished from merger-driven shocks and are not addressed further here.

Observational studies of merger shocks provide critical insights into merger dynamics and particle acceleration mechanisms. These approaches include detection of surface brightness and temperature discontinuities in X-ray data \citep{2002ApJ...567L..27M} and pressure profiles from Sunyaev-Zeldovich (SZ) \citep{1970Ap&SS...7....3S} observations. However, such measurements are challenging due to the low electron densities at cluster peripheries and instrumental sensitivity limits in X-ray and SZ bands. Merger shocks can accelerate electrons to relativistic energies while amplifying magnetic fields, producing detectable synchrotron emission in the radio frequencies. The resulting Mpc-sized, arc-like radio sources observed at cluster peripheries are termed radio relics \citep{2011ApJ...736L...8B}. Diffusive shock acceleration \citep{1983RPPh...46..973D} remains the leading proposed mechanism for electron acceleration at these shocks, though it has notable limitations in reproducing observed radio relic properties. The amplification of magnetic fields remains an open theoretical question. Additionally, ICM discontinuities can compress and re-accelerate fossil radio lobes from galaxies, generating phenomena known as radio phoenices \citep{2015MNRAS.448.2197D}.

In addition to radio relics, galaxy clusters host other large-scale diffuse radio sources associated with the ICM. Radio halos \citep{1999NewA....4..141G} predominantly occur in merging systems, though they have also been detected in relaxed clusters. On larger scales, megahalos \citep{cuciti+22} represent diffuse radio sources that encompass the radio halos themselves, producing a flattening of the exponential radio surface brightness profile at cluster outskirts \citep{2022SciA....8.7623B,2025ApJ...984L..26S}. We direct the reader to complementary chapters for detailed discussion of these sources \citep{Cuciti01.2026.SKA,Cassano01.2026.SKA}. 

The Square Kilometre Array Observatory (SKAO) will be the most sensitive large-scale radio telescope ever constructed, fundamentally transforming radio astronomy. This chapter examines the anticipated scientific impact of SKAO observations on radio relics and radio phoenixes, radio phenomena pertaining to shocks in merging galaxy clusters. The chapter is organised as follows: Sec.~\ref{shocks} introduces the types of shocks in cluster mergers, and Sec.~\ref{relics} reviews the use of radio relics as tracers of merger shocks. An overview of other shock-related radio sources is provided in Sec.~\ref{other}. The expected SKAO sensitivities for the various arrays are described in Sec.~\ref{ska-sens}. The open scientific issues that can be addressed with the SKAO, and the prospects for tackling them, are discussed in Sec.~\ref{ska-shocks}. Potential targets for science verification are identified in Sec.~\ref{sciverification}. Finally, the conclusions are presented at the end of Sec.~\ref{sciverification}.

\section{Shocks in cluster mergers} \label{shocks}

\begin{figure}
    \centering
    \includegraphics[]{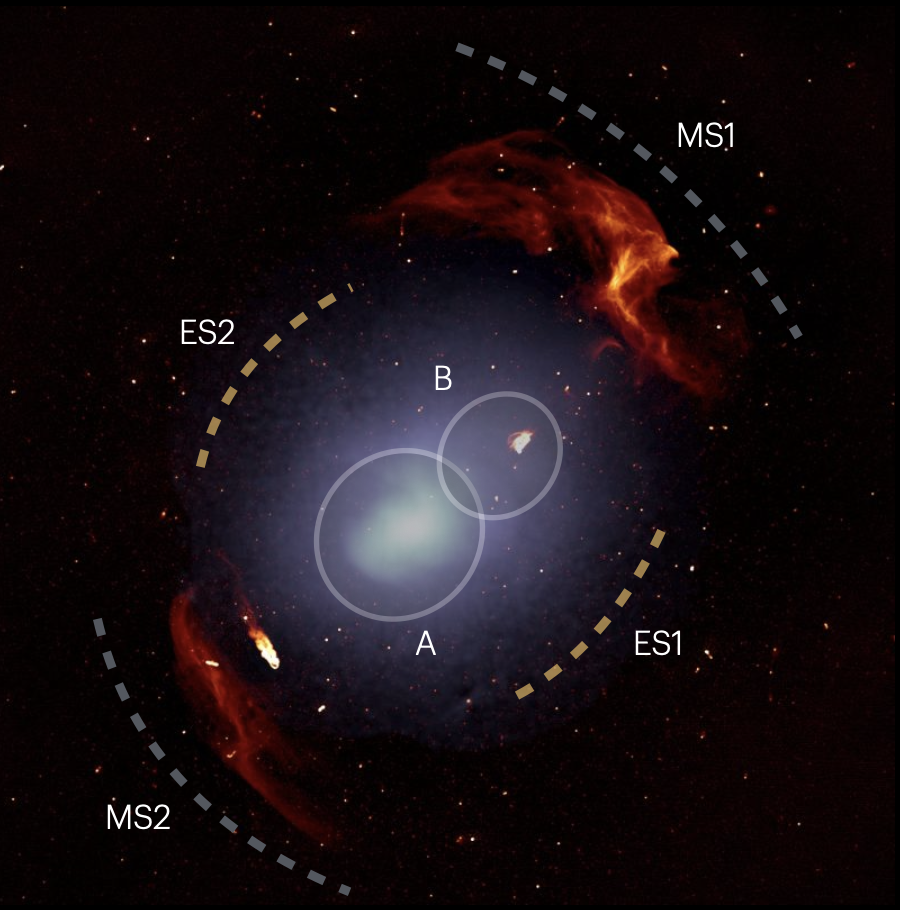}
    \caption{An example of a classic merger system in Abell 3667}, adapted from \citet{2022A&A...659A.146D}. The diffuse blue emission represents the XMM-Newton background-subtracted and exposure-corrected 0.5–2.0 keV image, with A and B indicating the two subclusters. The radio relics are shown in red. MS1 and MS2 mark the positions of the two merger shocks, while ES1 and ES2 indicate the expected locations of the equatorial shocks.
    \label{fig:schematic}
\end{figure}

Shock fronts represent one of two types of discontinuities observed in merging clusters, distinct from contact discontinuities (cold fronts). While cold fronts exhibit pressure equilibrium with density and temperature jumps of opposite sign, shock fronts are characterised by jumps in density and temperature of the same sign along with a pressure discontinuity \citep[e.g.][]{markevitch07}. The identification of shock fronts requires detection of a surface brightness edge in X-ray observations together with confirmation of the appropriate temperature jump.

The primary shock population in merging clusters consists of merger shocks (MS) directly induced by the collision of subclusters. In binary mergers, the collision generates a pair of merger shocks that propagate outward along the merger axis in opposite directions, commonly referred to as ``axial shocks'' \citep{Ha2018}. These are the most energetic shocks in merging clusters (MS1 and MS2 in Fig.~\ref{fig:schematic}). Cosmological simulations of head-on collisions with mass ratios $\sim$2 show that as merger shocks expand from the core to the outskirts, the average Mach number increases with time, with kinetic energy flux through the shocks peaking approximately 1~Gyr after their initial launching, at distances of $\sim$1--2~Mpc from the cluster centre \citep{Ha2018}. Due to the turbulent nature of the ICM, shock surfaces are not smooth but composed of many ``shock zones'' characterised by different Mach numbers. The Mach number distribution on merger shock surfaces can be approximated by a log-normal function, typically peaking at $\mathcal{M}_{\rm peak} \approx 2$--4.5 and extending up to $\sim$10 \citep{Lee24}. The Mach number can be estimated from the density or temperature jump using the Rankine--Hugoniot relations, assuming $\gamma = 5/3$ for a monatomic ideal gas:
\begin{equation}
\frac{T_2}{T_1} = \frac{(5\mathcal{M}^2 - 1)(\mathcal{M}^2 + 3)}{16\mathcal{M}^2}, \qquad \frac{\rho_2}{\rho_1} = \frac{4\mathcal{M}^2}{\mathcal{M}^2 + 3},
\end{equation}
where subscripts 1 and 2 denote upstream and downstream quantities respectively. X-ray observations with \textit{Chandra}, \textit{XMM-Newton}, and \textit{Suzaku} have detected merger shocks in a number of systems, including the Bullet Cluster \citep{2002ApJ...567L..27M} and Abell~2146 \citep{2010MNRAS.406.1721R}, typically revealing Mach numbers $\mathcal{M} \sim 1.5$--3. Detection remains challenging as shocks must be caught before moving to low surface brightness outer regions, and mergers must occur nearly in the plane of the sky to avoid projection effects. At radio wavelengths, merger shocks are believed to appear as radio relics.

At late times after pericenter passage, the infalling subcluster is decelerated by gravity while the bow shock continues propagating away from the cluster centre, creating a ``runaway'' merger shock detached from the driving subcluster \citep{2019MNRAS.488.5259Z}. Beyond $R_{500}$, steep gas density profiles ($\rho \propto r^{-3}$ or steeper) help maintain shock strength over large distances, creating a ``habitable zone'' in cluster outskirts where moderately strong shocks can be sustained or amplified. Runaway shocks are promising candidates for powering radio relics at cluster-centric distances of 2--3~Mpc through adiabatic compression of pre-existing relativistic particle populations from fossil AGN activity, producing boosted radio emissivity in narrow radial shells \citep{2019MNRAS.488.5259Z}.

Simulations also predict equatorial shocks (ES1 and ES2 in Fig.~\ref{fig:schematic}) that expand outward in the equatorial plane, perpendicular to the merger axis \citep{Ha2018}. These shocks form during the early compression phase as gas is squeezed perpendicular to the merger direction, typically launching before the axial shocks and propagating through low-density equatorial regions. Observational detections still remain rare, with candidates reported in only two systems to date \citep{Gu_2019NatAs, chen_2025ApJ}. Equatorial shocks are not discussed further in this chapter.

Accretion shocks form at or beyond the virial radius, between $\sim$0.9--1.5~$R_{200}$, as baryonic gas decelerates while falling into dark matter halos from the intergalactic medium \citep{Molnar2009}. They are characterised by high Mach numbers ($\mathcal{M} > 3$); the first SZ detection of a virial shock was reported in Abell~2319 at $(2.93 \pm 0.05) \times R_{500}$ with $\mathcal{M} > 3.25$ \citep{Hurier2019}. In terms of total energy dissipated in the ICM, accretion shocks are less significant than merger shocks, as they propagate at lower velocities into lower-density environments \citep{ryu2003}. At radio wavelengths, accretion shocks may contribute to faint extended emission features, though distinguishing these from merger-driven emission remains challenging \citep{hoeft07}. Since accretion shocks trace physics beyond merger-driven processes and are intimately connected to large-scale structure formation, a detailed discussion is presented in the chapter by \citet{Cuciti01.2026.SKA} As a possibility, runaway shocks can interact with accretion shocks forming merger-accelerated accretion shocks (MA-shocks), propagating way beyond the virial radius with potentially detectable signatures \citep{2020MNRAS.494.4539Z}.

\begin{figure}
    \centering
    \includegraphics[width=\textwidth]{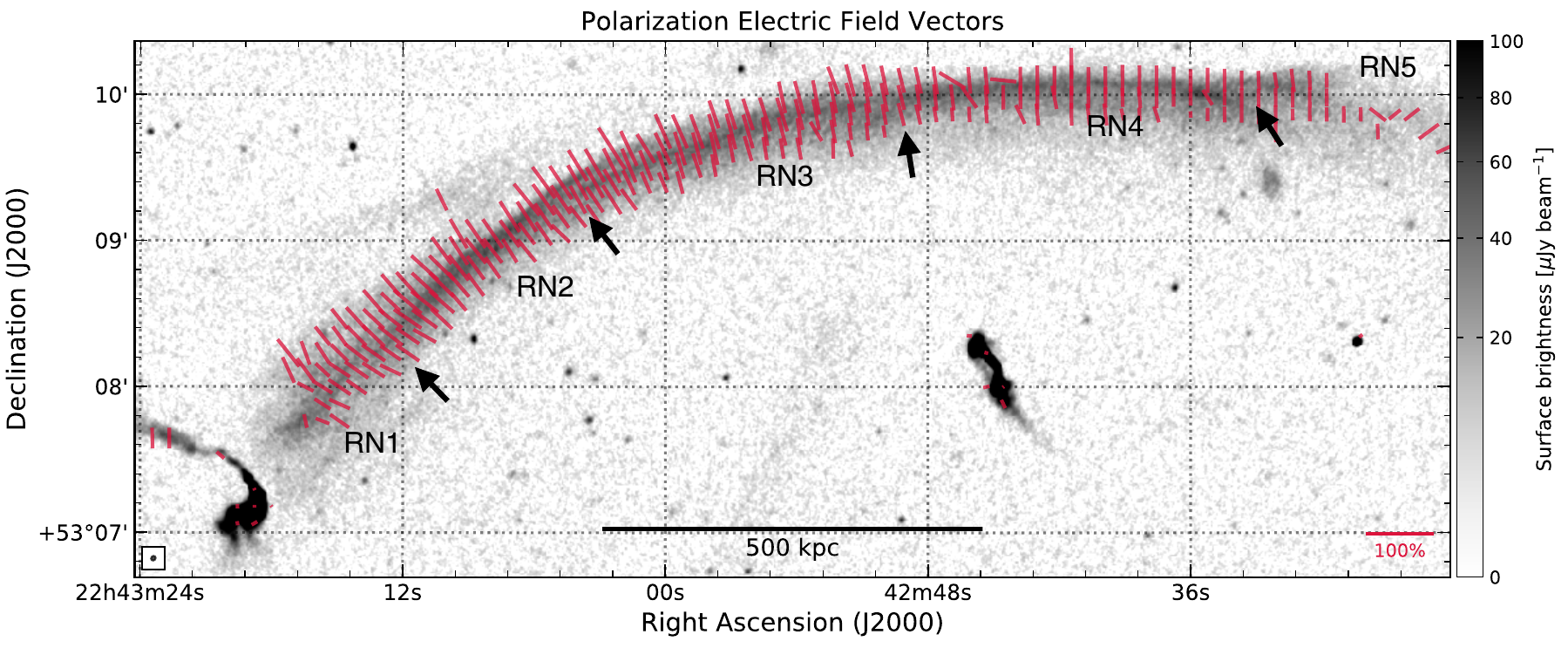}
    \includegraphics[width=1.05\textwidth]{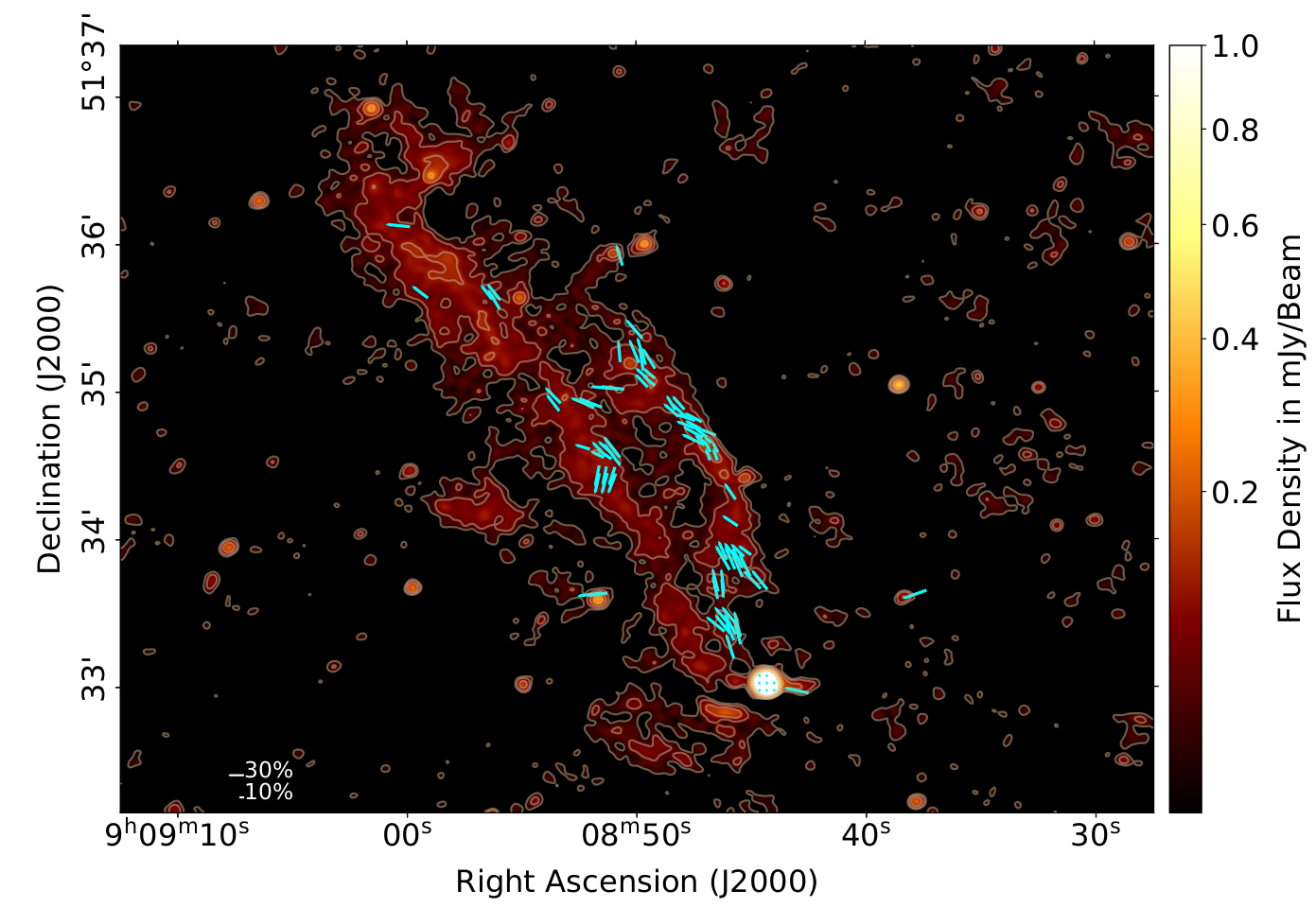}

    \caption{Top: High-resolution ($2.1'' \times 1.8''$) Stokes~I observation of CIZA\,J2242.8+5301 in the 1–2~GHz band, with polarization electric field vectors at $2.7''$ resolution shown in red \citep{digennaro+21b}. The vectors are corrected for Faraday rotation, and their lengths are proportional to the intrinsic polarization fraction (scale shown in the bottom-right corner). Black arrows in the two panels mark the locations where the relic breaks into separate filaments \citep{digennaro+21b}; Bottom: uGMRT 650~MHz intensity map of Abell 746 with a resolution of $4.6'' \times 4.2''$ shown in color scale. White contours start at $5\sigma$, with successive levels increasing by a factor of $\sqrt{2}$. Cyan vectors represent the magnetic field orientations (polarization angles rotated by $90^{\circ}$), corrected for the average RM, and their lengths are proportional to the fractional linear polarization. To illustrate the magnetic field distribution, the vectors are plotted after averaging over 4 pixels. \citep{2025ApJ...991..102P}}

    \label{fig:clusmerg}
\end{figure}

\section{Radio relics as tracers of shocks}\label{relics}
 Radio relics are believed to directly trace merger shocks propagating through the ICM. Understanding radio relics as shock tracers is crucial for studying particle acceleration at cluster shocks, probing the energetics of cluster mergers, and mapping magnetic field structures in cluster outskirts. Below we report some of the key features of radio relics. 

\subsection{Morphological and spatial connection to shocks}

The most compelling evidence that radio relics trace shocks comes from their morphology and location within merging clusters. Radio relics typically exhibit arc-like or elongated shapes located at projected distances of 0.5-2 Mpc from cluster centers, precisely where merger shocks are expected to propagate based on X-ray observations and numerical simulations \citep{Ha2018, Lee24}. The elongated morphology is naturally explained by the geometry of expanding shock fronts, with the radio emission tracing the shock surface as viewed in projection.

The surface brightness  morphology is consistent with particle acceleration occurring at the shock front, where electrons are injected to relativistic energies and emit the brightest synchrotron radiation. As particles age downstream of the shock, they lose energy through radiative cooling, producing progressively fainter and steeper-spectrum emission. High-resolution observations reveal that relics often show complex filamentary substructures on kpc scales \citep{2022A&A...659A.146D, 2022ApJ...927...80R}, potentially tracing inhomogeneities in the shock Mach number or pre-existing turbulent structure in the ICM.

In systems where both X-ray and radio observations are available, direct spatial coincidence between radio relics and X-ray surface brightness or temperature discontinuities has been established in several cases. For example, in CIZA J2242.8+5301 (the ``Sausage'' cluster), the prominent northern radio relic is located at the position of an X-ray shock with Mach number $\mathcal{M} \sim 2.7$ \citep{2015A&A...582A..87A}. In Abell 2146, deep X-ray observations revealed two shock fronts with Mach numbers $\mathcal{M} \sim 2$, and while initial radio observations at 325 MHz failed to detect extended emission, subsequent deep VLA observations at 1-2 GHz discovered extremely faint radio structures associated with both shocks \citep{2010MNRAS.406.1721R, 2018MNRAS.475.2743H}. These direct associations between X-ray shocks and radio emission provide strong evidence that radio relics are powered by merger shocks.

\subsection{Spectral signatures of shock acceleration}

The spectral properties of radio relics provide key diagnostic information about the underlying shock physics. Observations consistently reveal spectral index gradients across radio relics, with flat spectra ($\alpha \sim 0.6-1.0$, where the flux is $S_\nu \propto \nu^{-\alpha}$) at the outer edge steepening to $\alpha \sim 1.5-2.0$ toward the cluster center \citep{2012A&A...546A.124V, 2021A&A...646A..56R, 2025ApJ...979....4P}. This systematic gradient is a natural consequence of diffusive shock acceleration (DSA): freshly accelerated electrons at the shock front produce a power-law energy distribution with spectral index $\alpha_{\rm inj} \sim 0.5-0.8$, determined by the shock compression ratio. As particles advect downstream and age, they lose energy via synchrotron and inverse Compton radiation, causing the spectrum to steepen with distance from the shock \citep{1987PhR...154....1B}. However, \cite{jones+23} showed that the downstream width of the relics in the {\it Planck} clusters in LoTSS is very often greater than the theoretical expectations, which can be linked to downstream turbulence that further accelerates electrons in that region.

Under the DSA framework, the injection spectral index is related to the shock Mach number through the Rankine-Hugoniot jump conditions. For a strong shock with $\gamma = 5/3$, the spectral index is given by:
\begin{equation}
\alpha_{\rm inj} = \frac{1}{2}\left(\frac{r+2}{r-1}\right) - 1
\end{equation}
where $r = (\gamma+1)\mathcal{M}^2/((\gamma-1)\mathcal{M}^2+2)$ is the compression ratio. This relationship allows the shock Mach number to be inferred from radio observations: typical observed injection indices of $\alpha_{\rm inj} \sim 0.6-0.8$ correspond to $\mathcal{M} \sim 2.5-4$ \citep{2019SSRv..215...16V}. The integrated spectrum of radio relics, averaged over the entire relic volume, typically yields spectral indices $\alpha_{\rm int} \sim 1.0-1.3$, consistent with a mixture of freshly accelerated and aged electron populations.

\subsection{Polarization as a probe of shock-compressed magnetic fields}

Radio relics are among the most highly polarized extragalactic radio sources (Fig. \ref{fig:clusmerg}), with polarization fractions reaching 20-60\% at high frequencies ($\gtrsim$ 1.5 GHz) \citep[e.g.][]{digennaro+21b,2022A&A...659A.146D}. This high degree of polarization can be a direct consequence of magnetic field ordering by shock compression. As a shock wave passes through the magnetized ICM, the magnetic field component tangent to the shock surface is compressed and amplified by the shock compression ratio $r$. This compression creates an ordered magnetic field structure aligned parallel to the shock surface, leading to highly polarized synchrotron emission with electric vectors oriented perpendicular to the shock normal \citep{Ensslin1998}.

In the Toothbrush and Sausage relics, polarization fractions exceed 50\% at high frequencies with magnetic field vectors aligned with the elongated morphology of the relics \citep{2017A&A...600A..18K, 2012A&A...546A.124V,digennaro+21b}, exactly as expected if these structures trace expanding shock fronts. The polarization fraction is predicted to depend on the shock Mach number through the compression ratio, with stronger shocks producing higher polarization. For a Mach $\mathcal{M} = 3$ shock, the predicted intrinsic polarization fraction can reach $\sim$60-70\% for edge-on viewing geometries \citep{2022Galax..10...10H}, consistent with the highest values observed.

Recent observations have pushed polarization detections to lower frequencies (e.g. below 1 GHz), revealing that ordered magnetic fields persist even where Faraday depolarization becomes severe. In the merging cluster Abell 746, the northwest radio relic exhibits linear polarization of $\sim$18\% at 650 MHz, with polarization reaching $\sim$35\% at the shock's outer edge \citep{2025ApJ...991..102P}. The rotation measure-corrected magnetic field vectors remain well-aligned with the relic structure down to these frequencies, demonstrating substantial magnetic field coherence extending into the cluster outskirts. The detection of significant polarization at frequencies as low as 550 MHz shows that shock-compressed magnetic fields maintain their ordered structure across the emission region, even in the weaker field environment of cluster peripheries where shocks can more effectively modify and align the ambient fields \citep{2025ApJ...991..102P}.

\subsection{Energetic arguments and shock power}

The energetics of radio relics provide an independent line of evidence for their connection to merger shocks. The radio luminosity of a relic is directly related to the kinetic energy flux carried by the shock and the efficiency with which this energy is channeled into relativistic electrons. For a shock of surface area $A$ propagating with velocity $v_s$ through a medium of density $\rho$, the kinetic energy flux is $\Phi_{\rm kin} = \frac{1}{2}\rho v_s^3 A$. If a fraction $\xi_e$ of this energy is converted to relativistic electrons, the synchrotron luminosity depends on $\xi_e$, the magnetic field strength $B$, and the shock Mach number \citep{2007MNRAS.375...77H}.

Observations of radio relics show a strong correlation between radio luminosity and host cluster mass, $P_{1.4~{\rm GHz}} \propto M_{500}^{2-3}$ \citep{degasperin+14, Chatterjee_2024, Lee24,2025ApJ...979....4P}. 
The observed correlation provides quantitative support for the shock acceleration scenario and constrains the electron acceleration efficiency to be $\xi_e \sim 10^{-3}$ to $10^{-2}$ for typical magnetic field strengths of a few $\mu$G \citep{2020A&A...634A..64B}.

The total energy in relativistic electrons inferred from radio observations ($E_e \sim 10^{59-60}$ erg for typical relics) is consistent with the energy available from merger shocks given plausible acceleration efficiencies. Giant relics with sizes $> 2$ Mpc and radio powers $P_{1.4~{\rm GHz}} > 10^{25}$ W Hz$^{-1}$ are found exclusively in the most massive cluster mergers with $M_{500} \gtrsim 8 \times 10^{14}$ M$_{\odot}$ \citep{2024A&A...686A..55L}, where the available kinetic energy is sufficient to power such luminous radio sources. This mass-luminosity relationship and the energetic consistency between radio observations and merger shock models provide strong support for radio relics as direct tracers of merger shocks in galaxy clusters.

\section{Other shock-influenced radio sources}\label{other}
While radio relics are the most common tracers of cluster merger-generated shocks, recent observations have revealed a broader diversity of shock-influenced synchrotron sources in the ICM. These sources provide complementary insights into the microphysics of shocks, particle re-acceleration mechanisms, and the interplay between AGN activity and cluster-scale dynamics.

Radio phoenices are ultra-steep spectrum sources found within the cluster periphery and are believed to originate from aged plasma of radio galaxies that has been revived through compression and re-acceleration due to shocks in the surrounding medium. The adiabatic compression model by \citet{Ensslin2001A&A...366...26Eaddicomp} proposed revival of radio galaxy lobes via adiabatic compression by shocks in the ICM. These phoenixes are morphologically and spectrally different from the classical radio relics. They typically exhibit relatively low surface brightness, ultra–steep spectra ($\alpha < - 1.5$) with strong spectral curvature, indicative of aged electron populations and only mild polarization. Unlike the elongated, arc-like morphologies characteristic of relics tracing merger shocks, phoenices often display filamentary, patchy, or irregular structures, reflecting their complex evolutionary histories. With the advent of the SKA pathfinders (e.g. MeerKAT, uGMRT, LOFAR, ASKAP), the number of identified radio phoenices has grown rapidly, revealing a continuum of shock-influenced sources bridging the properties of classical relics and revived fossil plasma.

The well-studied case of Abell~85 demonstrates these features vividly: deep uGMRT and MeerKAT observations reveal a highly filamentary phoenix region with steep and curved spectra, consistent with the revival of fossil AGN plasma by a weak merger-driven compression or shock \citep{Rahaman2022MNRAS.515.2245RA85,Raja2023MNRAS.526L..70Rbridge,Raja2024ApJ...975..125Rphoenix}.

The combination of spectral curvature, spatial variation, and morphological correlation with X-ray substructures supports the fossil-plasma plus adiabatic-compression interpretation.

Complementary studies of other clusters also show that such revived fossil plasma regions may co-exist with or evolve into larger-scale relic-like features \citep{Rahaman2022MNRAS.509.5821RA1914}
, suggesting a possible evolutionary link between phoenixes and cluster shock relics. New theoretical developments further extend this picture: a hadronic-secondary model has been proposed in which relativistic protons produce secondary $e^{\pm}$ pairs that radiate in magnetized filaments, capable of reproducing both the steep/curved spectra and, in some cases, quasi-power-law behavior at higher frequencies \citep{Keshet2025arXiv250307714Kphoenix}.
This scenario, however, requires specific magnetic and pressure conditions and remains an active area of observational testing.

Radio galaxy tails sometimes exhibit spectral flattening features that are indicative of a process that rejuvenates the radio emission via compression or re-acceleration. These are distinct from phoenices as the radio jets from the core are sill active and can be traced to a host galaxy. Here we limit our scope to radio galaxies that are in galaxy clusters.

An example of this kind was discovered in the merging cluster Abell 1033 by \citet{deGasperin2017_1033} and has been referred to as "Gently Re-Energised Tail (GReET)". Radio galaxies with active cores embedded within clusters often exhibit distorted morphologies due to interactions with the surrounding ICM. However, some are distinguished by an unexpected increase in surface brightness at low radio frequencies, accompanied by a flattening of the spectral index in the trailing part of the tails. This peculiar behaviour is interpreted as evidence of a gentle re-energisation of aged relativistic electrons, which only barely compensates for radiative losses, likely driven by weak ICM shocks or low-level turbulence \citep{deGasperin2017_1033}. Other examples of radio galaxy tails showing such phenomena around clusters are head-tail galaxies in the clusters IIZW108 \citep{2021MNRAS.508.5326M} and ZwCl0634.1+47474 \cite{2024MNRAS.528..141L}.

Low-frequency observations of galaxy clusters have revealed a growing population of tailed radio galaxies exhibiting signatures of similar re-acceleration processes (e.g. \citealt{Pasini_2022, Lusetti_2024}). However, the origin of GReETs remains a subject of active investigation. The prevailing theory suggests that the re-energisation in GReETs is primarily driven by turbulent re-acceleration within the intracluster medium (ICM) \citep{deGasperin2017_1033,Lusetti_2024}. However, an origin linked to weak shocks propagating through the ICM cannot be ruled out and has been proposed for some systems \citep{Pasini_2022}.

\section{SKAO: AA* and AA4}\label{ska-sens}

The SKAO sensitivity calculators offer a range of combinations of the arrays that will be available. Starting from the AA2 for SKA-Low during the Science Verification, the array assemblies will go towards AA* and AA4. In this chapter, our interest is in the resolutions that these array assemblies will allow in order to probe the radio relics in exquisite detail.

\begin{figure}
    \centering
    \includegraphics[width=0.7\linewidth]{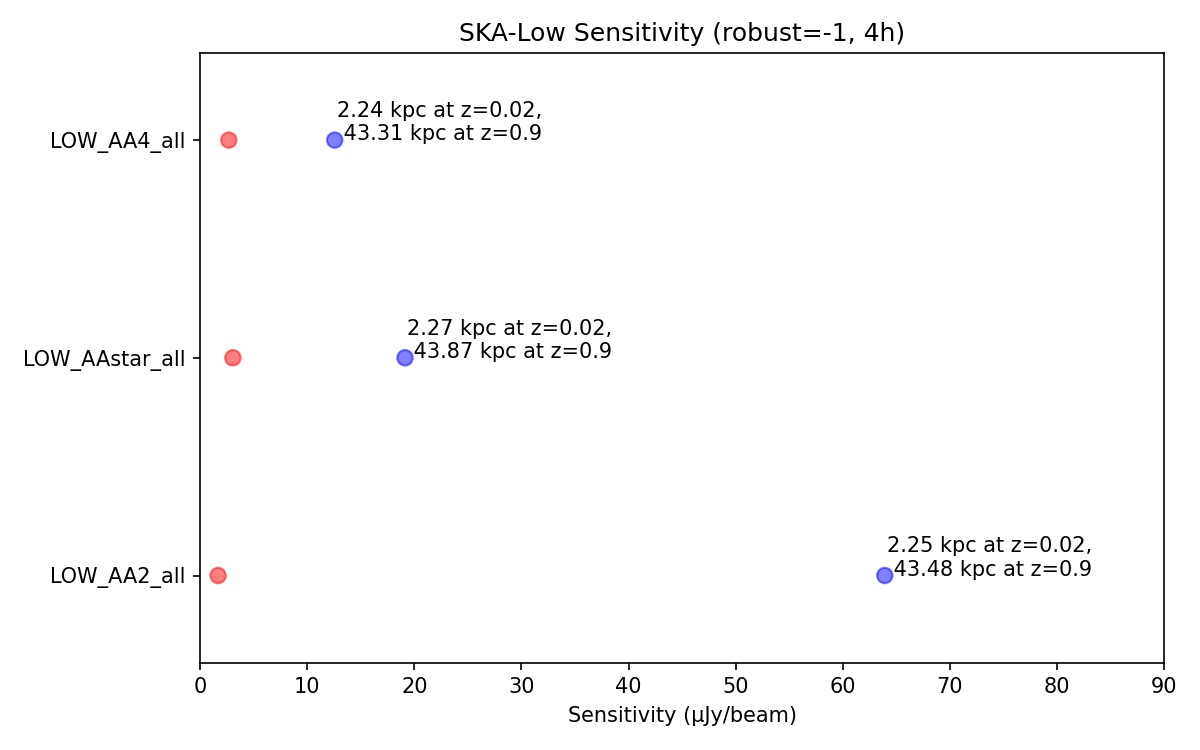}
    \caption{Comparison of sensitivities of the various SKAO arrays for SKA-Low. The sensitivities for 4 hours integration time are shown in blue and the confusion noise in red circles. The sizes of the circles scale relative to the major axes of the actual beam sizes that are $5.5''\times3.0''$, $5.5''\times4.7''$ and $5.5''\times4.3''$ for AA2, AA* and AA4, respectively. The annotated values in kpc correspond to the linear sizes that can be resolved at redshifts 0.02 and 0.9, respectively with the given resolution of the arrays AA2, AA* and AA4.}
    \label{fig:sens-ska-low}
\end{figure}

In Fig.~\ref{fig:sens-ska-low} we show the sensitivities for a range of arrays for an integration time of 4h, bandwidth of 75 MHz\footnote{A full bandwidth of 150 MHz may be available and will improve the sensitivity by $\sqrt{2}$. However the effects of combining this large bandwidth into a single image need to be understood.}, a central frequency of 200 MHz and robust$=-1$ weighting for visibilities. The time 4h was chosen to represent the expected observing time for a typical target in the SKA-Low AA2 Science Verification. In the context of the relics, the linear scales that can be resolved with the given resolution is of interest. We will be able to resolve at 2.3 kpc at redshift 0.02 and at 43.5 kpc at redshift 0.9, given the cosmology ($\mathrm{H}_0 =69.6$, $\Omega_M=0.286$ and $\Omega_{vac} = 0.714$).

\begin{figure}
    \centering
    \includegraphics[width=0.95\linewidth]{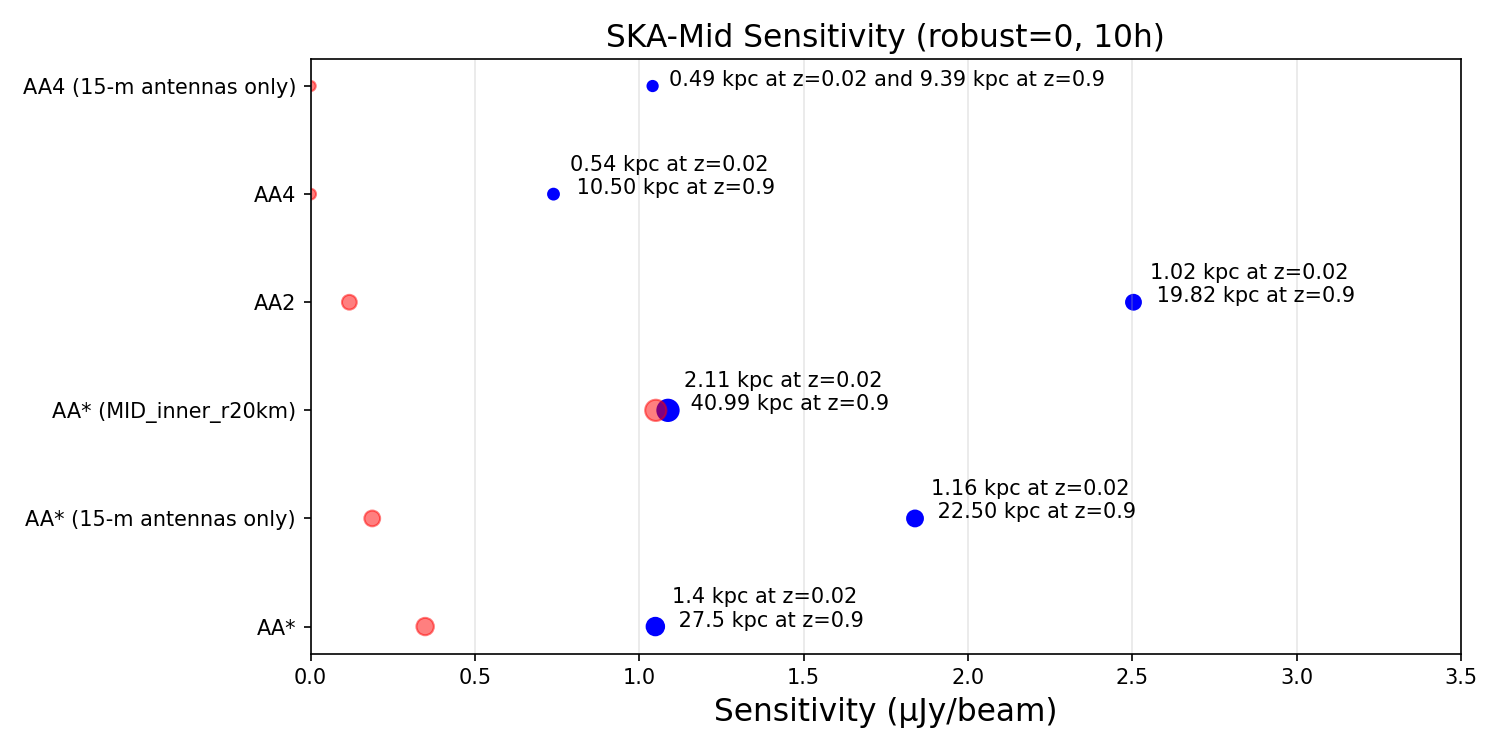}
    \caption{Comparison of sensitivities of the various SKAO arrays for SKA-Mid for which the confusion noise (red circles) is less than the sensitivity (blue) for a 10h observation. Much longer observations would be possible where the confusion noise is well below the 10h sensitivity. The sizes of the circles scale relative to the major axes of the actual beam sizes that are $3.5''\times3.0''$ (AA* ), $2.9''\times2.4''$ (AA* 15m antennas only), $5.2''\times4.4''$ (AA* mid inner r20km), $2.5''\times2.0''$ (AA2), $1.3''\times1.1''$ (AA4) and $1.2''\times1.0''$ (AA4 15m antennas only). 
    The labels next to the point indicate the linear sizes that can be probed for two representative redshifts of 0.02 and 0.9.}
    \label{fig:sens-ska-mid}
\end{figure}

In Fig.~\ref{fig:sens-ska-mid} we present a representation of the sensitivity for the Mid-arrays. The symbol sizes are scaled with the corresponding beam sizes. Here we have used an integration time of 10h, a central frequency of 0.79 GHz, bandwidth of 0.435 GHz and robust$=0$ weighting for the visibilities. 

The filamentary nature of the relics can be studied at these resolutions to understand the magnetic field configurations. Except the SKA-Mid configuration AA*(MID\_inner\_r20km), the confusion noise is well below the 10h continuum sensitivity. This will allow for longer integrations to reach the necessary sensitivities to image the relics with good significance at the corresponding resolution.

The upcoming SKA AA* and AA4 configurations will deliver $\mu$Jy-level sensitivity and arcsecond-scale resolution, enabling high dynamic range imaging of diffuse radio structures up to Mpc-scales. In particular, the extended mid-baseline coverage in the AA4 configuration reaching up to $\sim$159.6 km will significantly enhance angular resolution, achieving $\sim$3 arsec at 200 MHz and $\sim$6 arsec around 75 MHz, offering images comparable in resolution to that of the GMRT at 610 MHz and LOFAR at 150 MHz, but with improved uv-coverage. This will allow detailed studies of the relics and low surface brightness emission around them. The high-resolution, multifrequency observations will further enable precise mapping of spectral steepening and aging behind cluster shocks, allowing clear separation of shock injection sites, compression regions, and turbulent structures within the intracluster medium (ICM). The enhanced sensitivity at low frequencies will also allow detection of relics in low mass clusters as well as in faint populations at higher redshifts, $z\geq $1. With the enhanced survey speed of the SKA, it will be possible to study a statistical sample of relic power with respect to cluster mass as well as Mach number and provide a better constraint on large-scale structure formation and evolution in the Universe. In addition, AA2 will provide opportunities for science verification on well-known bright relics.

\section{Addressing open questions with the SKAO}\label{ska-shocks}
The open scientific issues pertaining to radio relics and merger shocks span both the microphysics of particle acceleration and magnetic field amplification in individual systems, and the occurrence and evolution of these phenomena in the cosmological context. In this section we identify the key open questions and outline how the SKAO capabilities described in the previous section will advance our understanding.

\subsection{Magnetic field distribution, amplification, and depolarization}
\label{sec:Bfield}

The origin and strength of magnetic fields in radio relics represent one of the most pressing open problems. Equipartition estimates and Faraday rotation measure (RM) analyses consistently yield field strengths of $\sim$1--5~$\mu$G in relics \citep{2017A&A...600A..18K, digennaro+21b}, comparable to values found in the central ICM despite the much lower gas density at the cluster periphery where relics reside. Standard adiabatic shock compression amplifies the pre-shock magnetic field component parallel to the shock surface by the compression ratio $r$, yielding an enhancement of only a factor of $\sim$2--4 for typical merger shocks with $\mathcal{M} \sim 2$--3 \citep{2012MNRAS.423.2781I}. If upstream fields in cluster outskirts are $\sim$0.1--1~$\mu$G, as suggested by RM studies of background radio galaxies and cosmological simulations, compression alone is insufficient to account for the observed field. As alternatives, cosmic ray-driven streaming instabilities in the shock precursor can amplify the upstream field by factors of 10--20 before the gas reaches the shock \citep{2004MNRAS.353..550B, 2013MNRAS.436..294B}. Also, turbulent dynamo action in the post-shock region, seeded by vorticity generated at the shock discontinuity, can further amplify fields on eddy turnover timescales \citep{2019ApJ...883..138R} and baroclinic vorticity generation arising from misalignment between pressure and density gradients in an inhomogeneous ICM provides an additional source of turbulent energy for field growth \citep{2012MNRAS.423.2781I}. The relative importance of these processes with their dependence on the shock Mach number, upstream field geometry, and ICM thermodynamic conditions, remains poorly constrained.

High-resolution observations have added another layer of complexity by revealing pervasive filamentary substructures on scales of a few to tens of kpc in numerous relics, including the Sausage relic \citep{digennaro+21b}, the Toothbrush relic \citep{2017ApJ...835..197V, 2022ApJ...927...80R}, Abell~3667 \citep{2022A&A...659A.146D}, and MACS~J0717.5+3745 \citep{2022ApJ...934...49G}. The origin of these filaments is still debated. they may reflect spatial variations in the shock Mach number across a rippled shock surface, producing localised regions of enhanced particle acceleration \citep{2013ApJ...765...21S}, magnetic reconnection between current sheets at boundaries of magnetic domains of different orientation \citep{2022ApJ...927...80R}, re-acceleration of pre-existing fossil plasma clouds whose spatial distribution imprints a filamentary morphology \citep{2021MNRAS.500..795D}, or fluctuations in magnetic field strength and topology that concentrate synchrotron emissivity into regions of locally enhanced field \citep{2014ApJ...794...24O}. Work on Abell 3667 suggests that most of the magnetic energy is concentrated within the filaments themselves, while inter-filament regions have field strengths comparable to the unperturbed ICM at the same cluster-centric distance \citep{2022A&A...659A.146D}, implying that filaments are genuinely distinct magnetic structures, possibly magnetic flux tubes through which relativistic particles stream at high velocity. Distinguishing between these scenarios requires spatially resolved spectral index mapping across individual filaments and Faraday rotation studies to probe the magnetic field coherence length and power spectrum, observations that remain at the limit of current facilities.

Polarimetric observations offer an additional view of the magnetic field structure that total intensity measurements cannot provide. Radio relics are among the most polarized extragalactic radio sources, with polarization fractions reaching 20\% to 60\% at GHz frequencies \citep{digennaro+21b, 2022A&A...659A.146D}. This aligns with the significant field ordering caused by shock compression. However, the magnetic field also depolarizes the signal through various distinct processes \citep{1966MNRAS.133...67B, 1998MNRAS.299..189S}. External Faraday dispersion (EFD) happens when a non-emitting magnetized screen lies in front of the source. Turbulent cells within the telescope beam create differential Faraday rotation, which is described by $p/p_0 = e^{-2\sigma_{\rm RM}^2\lambda^4}$. Internal Faraday dispersion (IFD) occurs when the emitting and Faraday-rotating regions overlap, causing the polarization plane to take a random path through the emission volume, expressed as $p/p_0 = (1-e^{-2\sigma_{\rm RM}^2\lambda^4})/(2\sigma_{\rm RM}^2\lambda^4)$. In both cases, $\sigma_{\rm RM}$ represents the RM dispersion and $p_0$ is the intrinsic polarization fraction \citep{1966MNRAS.133...67B}. Additional wavelength-independent depolarization can result from unresolved structure in the intrinsic polarization angle distribution or from complex three-dimensional projection effects.

Disentangling these mechanisms observationally has proven difficult. Detailed 1--4~GHz \textit{QU}-fitting of the Sausage relic revealed decreasing polarization in the downstream region regardless of the assumed depolarization model, explained only by geometrical projections combined with increasing magnetic field anisotropy toward the cluster centre \citep{digennaro+21b, dominguez-fernandez+21}. In the Toothbrush relic, strong wavelength-dependent depolarization between 1.4 and 18.6 GHz indicates substantial Faraday dispersion linked to the relic's embedding within the ICM \citep{rajpurohit+20}. In PSZ2 G096.88+24.18, the southern relic shows depolarization consistent with EFD from turbulent ICM with coherence scales of $\Lambda_{\rm min} \sim 35$~kpc to $\Lambda_{\rm max} \sim 400$~kpc, while the northern relic exhibits much higher polarization, suggesting less turbulent surroundings or a different viewing geometry \citep{2024A&A...691A..23D}. In other systems, IFD has been preferred \citep{stuardi+19, 2022A&A...666A...8S}, implying a turbulent ICM upstream of the shock. More recently, single-component models have been shown to fail entirely in some cases: in Abell~746, the northwest relic exhibits $\sim$18\% linear polarization at 650~MHz with polarization reaching $\sim$35\% at the shock's outer edge, but requires two-component depolarization models combining multiple emission regions along the line of sight with fitted RM dispersions of $\sigma_{\rm RM,1} \sim 30$--36~rad\,m$^{-2}$ for the far-side component and $\sigma_{\rm RM,2} \sim 2$--3~rad\,m$^{-2}$ for the near side \citep{2025ApJ...991..102P}. Similarly, polarization studies of MACS\,J0717.6+3745 up to 10~GHz reveal a frequency-dependent change in the magneto-ionic structure \citep{pasetto+25}. The detection of significant polarization at 550--750~MHz in Abell~746 overturns the earlier paradigm that relics are completely depolarised below 1~GHz and demonstrates that depolarization can proceed in a step-wise rather than smooth manner, revealing magnetoionic complexity that existing observations cannot fully characterise.

Simulations predict non-Gaussian rotation measure (RM) distributions in relics because of the compression of turbulent magnetic field structure in the emitting region \citep{2018MNRAS.474.1672V, 2019MNRAS.490.3987W, dominguez-fernandez+21}. While some observations support this prediction \citep{2022A&A...666A...8S}, others find more Gaussian distributions \citep{digennaro+21b}. High-resolution studies reveal significant spatial RM variations on scales of 15--30~kpc \citep{digennaro+21b, 2024A&A...691A..23D}, indicating complex small-scale structures. A fundamental limitation of current observations is that polarization measurements at different frequencies often sample different spatial scales due to resolution differences, and lack the sensitivity to detect low surface brightness polarized emission at the lowest frequencies where depolarization effects are most severe and most diagnostic.

SKAO's wideband spectro-polarimetric capabilities will be essentia to finally understand these systems. Detailed RM synthesis across SKA-Low (50--350~MHz) and SKA-Mid Band~1--2 (0.35--1.76~GHz) will map the spatial distribution and power spectrum of magnetic field fluctuations at resolutions of a few kpc in nearby systems for the first time. By constructing RM structure functions and comparing them with MHD simulations of shock-amplified turbulence \citep{dominguez-fernandez+21}, it will become possible to measure the outer scale and spectral slope of the magnetic turbulence, distinguishing cosmic ray-driven instabilities (which predict enhanced power on scales of $\sim$10--100~kpc set by the diffusion length) from turbulent dynamo processes (which predict a Kolmogorov-like spectrum extending to smaller scales). Independent field strength estimates from synchrotron spectral break frequencies, measured with SKA-Mid Band~5 (4.6--15.3~GHz), will provide spatially resolved maps of field strength across relic faces that can be cross-checked against equipartition and RM-based values.

With the exceptional resolution and sensitivity, SKAO is expected to resolve individual filaments in relics out to $z \sim 0.3$--0.5 and map spectral indices across and along them. If filaments are discrete acceleration sites at a rippled shock, spectra should be flat within filaments and steepen at their boundaries. While in the case of magnetic flux tubes, spectra should vary smoothly along the field-aligned direction. High intrinsic polarization within filaments but lower polarization in inter-filament regions would strongly support the flux tube interpretation.

The continuous frequency coverage from 50~MHz to 1.76~GHz spans the $\lambda^2$ range ($\sim$0.03--36~m$^2$) where both IFD and EFD produce their most distinctive signatures. The RM resolution achievable with Faraday synthesis across the full SKA-Low band is $\delta\phi \approx 1$~rad\,m$^{-2}$, sufficient to resolve the Faraday-thin components in current multi-component models and to search for additional complexity. Crucially, SKAO will observe with comparable angular resolution across the full band, eliminating the systematic uncertainty from resolution-dependent sampling that plagues current multi-frequency polarization studies. A statistical sample of relics spanning a range of Mach numbers, cluster masses, redshifts, and merger geometries will establish whether the observed diversity in (de)-polarization behaviour reflects genuine physical differences in upstream ICM turbulence and shock properties or arises primarily from viewing geometry, test the predicted non-Gaussianity of RM distributions, and map how the RM power spectrum evolves from the upstream to the downstream side of the shock.

\subsection{Particle acceleration efficiency and seed electrons}
\label{sec:seed}

A key challenge for the diffusive shock acceleration framework is the so-called efficiency problem. The kinetic energy dissipated at the shock is generally insufficient to reproduce the observed radio luminosities if electrons are drawn purely from the thermal pool \citep{2005ApJ...627..733M, 2020A&A...634A..64B}. The problem is most acute for relics associated with weak shocks ($\mathcal{M} \lesssim 2.5$), where DSA theory predicts steep electron energy spectra and correspondingly fainter emissions.

The leading resolution to this problem invokes re-acceleration of a pre-existing population of mildly relativistic ``fossil'' electrons with Lorentz factors $\gamma_e \sim 10^2$--$10^4$, likely injected by previous AGN activity, earlier structure formation shocks, or turbulent acceleration in the ICM \citep{2013MNRAS.435.1061P, 2016ApJ...823...13K}. Fossil electrons enhance the effective acceleration efficiency by providing seed particles already above the thermal injection threshold, bypassing the bottleneck of extracting electrons from the thermal pool. This scenario can simultaneously explain the high radio luminosities of some relics and the low overall occurrence rate of radio relics ($\sim$10\% of merging clusters; \citealt{jones+23}): shocks may only produce bright relics when they encounter clouds of fossil electrons, while shocks propagating through ``clean'' ICM remain radio-quiet. Direct observational evidence linking specific relics to fossil AGN plasma has been found in a handful of systems, most notably the complex of tailed radio galaxies feeding the relic in Abell~3411--3412 \citep{vanweeren+17} and the connection between an AGN and the relic in CIZA~J2242.8+5301 \citep{bonafede+14}. However, the ubiquity of fossil electron populations in cluster outskirts, their spatial distribution, energy spectrum, and volume-filling fraction remain essentially unconstrained.

A closely related puzzle concerns the electron injection problem at weak shocks. Particle-in-cell simulations suggest that quasi-perpendicular shocks with $\mathcal{M} \lesssim 2.3$ may be unable to inject thermal electrons into the DSA cycle efficiently \citep{2016ApJ...823...13K}, yet many relics are associated with shocks in precisely this Mach number range. Alternative injection mechanisms have been proposed, including shock drift acceleration at oblique shocks \citep{2014ApJ...797...47G}, but their efficiency in the high-$\beta$ ($\beta \sim 100$) plasma characteristic of cluster outskirts is poorly understood. Recent work has explored multi-shock scenarios in which particles passing through successive weak shocks are progressively boosted to higher energies, significantly increasing the effective acceleration efficiency \citep{2023MNRAS.526.4234S}, but observational tests of such models remain limited.

Resolving the seed electron question requires two complementary capabilities that SKAO will provide. First, SKA-Low will be uniquely suited to search for fossil plasma populations in cluster outskirts. Remnant lobes and tails from radio galaxies develop ultra-steep spectra ($\alpha \lesssim -2$) as their electron populations age through synchrotron and inverse Compton losses, rendering them invisible above $\sim$1~GHz within a few hundred Myr \citep{2020A&A...634A...4M, 2024Galax..12...19V}. At SKA-Low frequencies (50--350~MHz), such aged plasma remains detectable for up to $\sim$0.5--1~Gyr after injection, depending on the local magnetic field strength and ICM environment \citep{2024Galax..12...19V}. The approximately eight-fold improvement in sensitivity over LOFAR \citep{2019arXiv191212699B} will allow SKA-Low to reach surface brightness levels of order 0.1~$\mu$Jy\,arcsec$^{-2}$ in deep integrations, probing fossil electron populations with $\gamma \sim 10^2$--$10^3$---well into the energy range relevant for re-acceleration by weak merger shocks \citep{2014IJMPD..2330007B}. Systematic ultra-low-frequency mapping of the upstream regions of known relics will establish whether fossil plasma is ubiquitous in the shock environment or confined to specific locations where AGN have recently been active, directly testing the seed electron hypothesis.

Second, SKA-Mid's broad frequency coverage (0.35--15.4~GHz) will enable spatially resolved measurement of the injection spectral index $\alpha_{\rm inj}$ at the shock front and its variation along the relic length. In the DSA framework, $\alpha_{\rm inj}$ is set by the shock compression ratio and hence by the Mach number; any systematic offset between the radio-derived Mach number (from $\alpha_{\rm inj}$) and the X-ray-derived Mach number (from the density or temperature jump) provides a direct diagnostic of re-acceleration of pre-existing electrons, since fossil seed particles modify the post-shock energy spectrum relative to thermal-pool injection \citep{2020A&A...634A..64B}. Combined SKA-Low and SKA-Mid observations will provide continuous spectral coverage from 50~MHz to beyond 5~GHz, spanning the frequency range over which spectral curvature, the telltale signature of re-acceleration as opposed to fresh injection, becomes most diagnostic \citep{2022MNRAS.509.1160I}.

\subsection{Upstream and downstream physics of the shock}
\label{sec:updown}

The regions immediately upstream and downstream of the shock front hold key information about collisionless shock acceleration, yet both remain poorly characterised. Upstream, the potential existence of a synchrotron precursor powered by cosmic ray-driven instabilities would directly reveal how particles are injected and how magnetic fields are amplified before the gas is compressed. Downstream, the spectral profile of the ageing electron population encodes the post-shock magnetic field evolution, the advection velocity, and the possible role of secondary re-acceleration. Together, these two regions bracket the shock and define the boundary conditions for any complete model of particle acceleration at merger shocks.

In supernova remnants, streaming cosmic ray protons excite resonant and non-resonant instabilities that amplify upstream magnetic fields by orders of magnitude \citep{2004MNRAS.353..550B}. If analogous processes operate at weaker merger shocks, pre-accelerated electrons diffusing ahead of the shock could produce detectable synchrotron radiation in the amplified field \citep{2013MNRAS.436..294B}. For a $\mathcal{M} \sim 2$--3 shock under typical cluster outskirt conditions, the predicted precursor extends $\sim$50--100~kpc ahead of the shock with surface brightness $\sim$0.5--2~$\mu$Jy\,arcsec$^{-2}$ at 150~MHz \citep{2013MNRAS.436..294B}. X-ray precursor signatures have been tentatively identified in supernova remnants \citep{2014ApJ...781...65W}, but no equivalent radio detection exists for ICM shocks, though one candidate has recently been reported at very low frequencies \citep{2025A&A...699A.200L}. The non-detection may reflect less efficient field amplification at weak shocks, intrinsically low upstream fields, or projection effects obscuring faint signals. Importantly, standard DSA and superdiffusive shock acceleration predict distinct upstream spatial profiles, exponential versus power-law decay of the particle density \citep{2017A&A...601A..64Z, 2018MNRAS.478.4922Z}, but distinguishing between them requires both high spatial resolution to isolate the shock edge and sufficient multi-frequency sensitivity in the faint upstream region.

On the downstream side, the simplest DSA picture predicts that electrons cool passively via synchrotron and inverse Compton losses while advecting away from the shock, producing characteristic radio widths of $\sim$50--100~kpc at GHz frequencies \citep{2007MNRAS.375...77H} with progressive spectral steepening behind the shock front. This qualitative trend is confirmed in several well-studied relics \citep{2018ApJ...852...65R, 2022A&A...659A.146D, 2025ApJ...979....4P}, but quantitative comparisons reveal systematic discrepancies. The Toothbrush relic shows emission extending $\sim$800~kpc downstream at 58~MHz \citep{degasperin20toothbrush}, far exceeding predictions even accounting for projection. Spectral age modelling of the Sausage relic yields electron ages systematically younger than expected from the inferred shock speed and downstream distance \citep{2014MNRAS.445.1213S, digennaro+18, 2025ApJ...978..122L}. More broadly, the downstream widths of relics in the \textit{Planck} clusters observed in LoTSS are very often greater than theoretical expectations \citep{jones+23}, pointing to a systematic failure of the laminar advection-plus-cooling model. Proposed explanations include ongoing turbulent re-acceleration in the post-shock region \citep{Kang2017, degasperin20toothbrush}, super-Alfv\'enic particle diffusion \citep{2014MNRAS.445.1213S}, complex three-dimensional shock geometry confusing age--distance relationships \citep{2019MNRAS.490.3987W}, episodic re-energisation by secondary shocks \citep{2023MNRAS.526.4234S}, and Rayleigh--Taylor instabilities triggered by upstream density fluctuations that generate downstream velocity turbulence \citep{2020ApJ...895..143N}. Which of these processes dominates remains an open question.

On the upstream side, in the most favourable cases where streaming instabilities amplify the field to $\sim$1~$\mu$G, the precursor surface brightness at 150~MHz reaches $\sim$0.01--0.1~$\mu$Jy\,arcsec$^{-2}$, potentially within reach of deep ($\gtrsim$100~h) SKA-Low observations tapered to $\sim$100~kpc scales. Even non-detections will place meaningful upper limits on the upstream field strength and cosmic ray diffusion coefficient. SKA-Mid Band~2 at $\sim$1$^{\prime\prime}${} resolution combined with matched SKA-Low imaging will measure the spectral index gradient across the shock edge, testing the standard DSA exponential profile against superdiffusive power-law predictions \citep{2018MNRAS.478.4922Z}.

On the downstream side, SKA-Mid Band~2 at 1--2$^{\prime\prime}$ond  will resolve the cooling region in nearby relics ($z \lesssim 0.1$) on kpc scales, while sub-arcsecond resolution will be essential at $z \sim 0.3$--0.5 where the cooling region subtends only a few arcseconds. The broad frequency coverage from SKA-Low through SKA-Mid, spanning from 50~MHz to beyond 5~GHz, is critical because different ageing models, Jaffe--Perola, Kardashev--Pacholczyk, and continuous-injection variants, predict distinct curvature profiles best distinguished over a wide baseline \citep{2022ApJ...927...80R, 2022MNRAS.509.1160I}. Spectral flattening in the far downstream or the persistence of emission at unexpectedly high frequencies would provide direct evidence for post-shock turbulent re-acceleration \citep{2011ApJ...728...82M, 2015ApJ...815..116F}. Simultaneously, SKA-Low at 50--200~MHz will detect the oldest downstream electrons that have cooled below current detection thresholds, extending the observable downstream region to its full physical extent \citep{2014IJMPD..2330007B}.

\subsection{Connection between radio relics and radio halos}
\label{sec:relicshalo}

Radio relics and radio halos are often observed in the same merging cluster, where relics trace shock acceleration at the periphery and halos trace turbulent re-acceleration in the cluster volume. The co-occurrence of these sources in systems such as Abell~2744, MACS~J0717.5+3745, the Coma cluster, and El Gordo points to a common origin in the merger energetics, yet the fraction of merger energy channelled into shocks versus turbulence and its dependence on the merger stage, mass ratio, and impact parameter remain poorly constrained. Notably, radio halos are detected in up to $\sim$20-40\% of the most massive merging clusters \citep{2013A&A...557A..99K,2021A&A...647A..50C}, whereas radio relics are found in only $\sim$10\% \citep{jones+23}. The physical origin of this difference in occurrence rates is not understood.

In a growing number of systems, faint bridges of diffuse emission connecting relics to halos have been reported \citep{2013A&A...551A.141M, 2022A&A...657A..56K}, hinting at physical continuity between the shock-accelerated and turbulence-accelerated electron populations. These bridges may represent transition regions where shock-accelerated particles are further processed by turbulence as they advect toward the cluster centre. Mapping the spectral index and polarization properties across the relic--bridge--halo transition would directly constrain how particle acceleration evolves from the shock-dominated to the turbulence-dominated regime, but current facilities lack the sensitivity to characterise this faint emission at adequate resolution.

SKAO's sensitivity to ultra-faint extended emission will enable spectral and polarimetric mapping across the relic--bridge--halo transition for the first time. Spectral steepening from relic into bridge and halo regions would trace how aged particles diffuse from the shock into the turbulent volume, while polarimetry will map the transition from the highly ordered shock-compressed field (20--60\% polarization) to the more turbulent halo field structure. RM synthesis across this boundary will probe the evolution of the magnetised plasma through the shock-to-turbulence transition. On the statistical side, SKAO's ability to detect halos in systems with marginal or absent relic emission---and vice versa---will establish correlations between relic and halo properties across diverse merger states, constraining the energy partition between shock and turbulent acceleration and testing whether the occurrence rate asymmetry is physical or observational in origin.


\subsection{Runaway shocks and distant relic populations}
\label{sec:runaway}

After pericenter passage in a merger, the infalling subcluster decelerates while the bow shock continues propagating outward, creating a ``runaway'' shock that detaches from the driving subcluster and travels to large cluster-centric distances beyond $R_{500}$ \citep{2019MNRAS.488.5259Z}. In the steep density profiles of cluster outskirts ($\rho \propto r^{-3}$ or steeper), these shocks can be sustained or even amplified, creating a ``habitable zone'' where moderately strong shocks persist over Mpc-scale distances. Runaway shocks are promising candidates for powering radio relics at distances of 2--3~Mpc from the cluster centre through adiabatic compression of pre-existing fossil AGN plasma in narrow shells behind the shock front \citep{2019MNRAS.488.5259Z}. When these shocks interact with the virial accretion shock, they can form merger-accelerated accretion shocks that propagate beyond the virial radius \citep{2020MNRAS.494.4539Z}. Observational characterisation of runaway shocks remains limited: the relics they power are expected to be faint and steep-spectrum, placing them below current detection thresholds at most frequencies.

SKA-Low's sensitivity at 50--350~MHz is well matched to the steep spectra expected from runaway-shock-powered relics. Deep observations of massive merging clusters will search for faint extended emission at 2--3~Mpc from the cluster centre, probing the full radial extent of shock propagation. Multi-frequency coverage across Low and Mid will map the spectral steepening with distance from the shock, constraining magnetic field strengths and diffusion coefficients in the rarefied cluster periphery. Polarimetry will test whether shock-compressed fields maintain ordered structure in these extremely low-density environments. Where runaway shocks interact with accretion shocks, SKAO may detect distinctive radio signatures at the collision interface, providing constraints on the multi-shock structure and cumulative energetics of cluster assembly.


\subsection{Population studies}
\label{sec:population}

\begin{figure}
\centering
\includegraphics[width=0.48\textwidth]{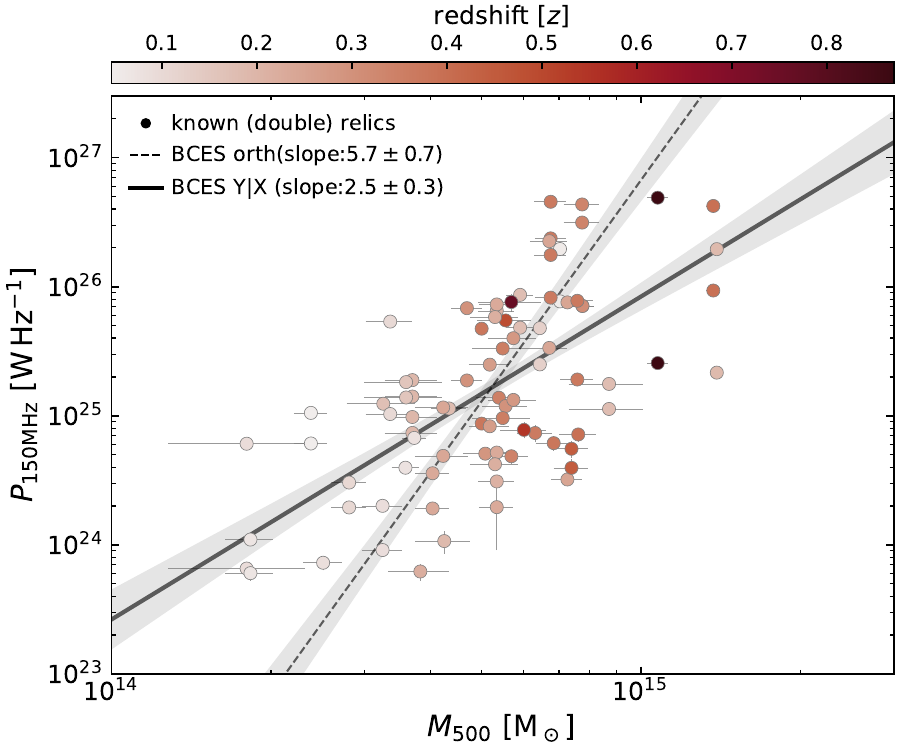}
\includegraphics[width=0.48\textwidth]{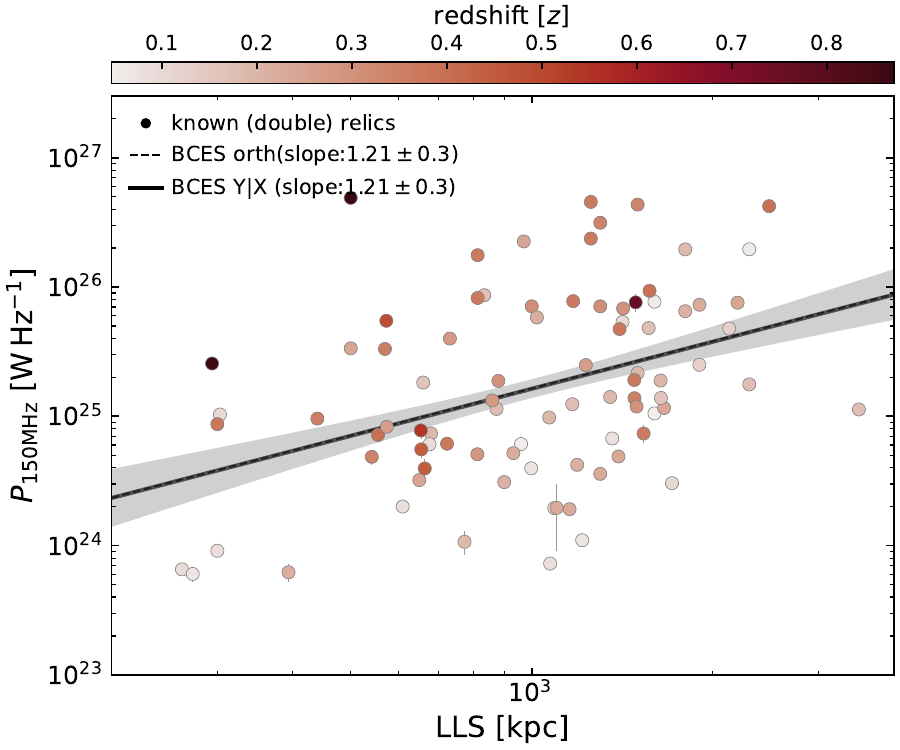}
\caption{Scaling relations for single/multiple radio relics. Left panel: radio power vs mass. Right panel: radio power vs largest linear size. Data points are taken from \cite{jones+23} and \cite{stroe+25}.}
\label{fig:correlations}
\end{figure}

At present, only a few tens of clusters hosting radio relics are known, and most detailed studies are performed on individual systems. Existing observations suggest correlations between relic radio power ($P_\nu$), largest linear size (LLS), and host cluster mass ($M_{500}$) related to the merger energetics and evolutionary stage \citep[Fig.~\ref{fig:correlations}]{degasperin+14, Chatterjee_2024, stroe+25}, but these are hampered by small-number statistics, Malmquist bias, and restriction to the local Universe and the most massive clusters. At high redshift, misclassifications with radio galaxies due to limited angular resolution is an additional concern. Comparison with simulations has been challenging: cosmological simulations systematically over-predict the number of relics, especially at high redshift \citep{nuza+12, nuza+17, wittor+21}. Recent TNG-Cluster simulations \citep{Lee24} confirm that the relic fraction increases with cluster mass, but still predict more high-$z$ relics than current LOFAR and MeerKAT surveys detect \citep{digennaro+21, jones+23, kolokythas+25}, suggesting either a large population of low-power relics below current sensitivity limits or a genuine shortcoming in the modelling.

SKAO will address the relic rarity problem through its combination of survey speed and sensitivity. At a sensitivity of $\sim$20~$\mu$Jy\,beam$^{-1}$ (10$^{\prime\prime}${} resolution), SKA-Low increases the expected number of detectable low-power relics at high redshift by an order of magnitude compared with LOFAR ($\sim$200~$\mu$Jy\,beam$^{-1}$ at similar resolution), reaching a minimum detectable radio power of $\sim$$10^{24}$~W\,Hz$^{-1}$ at $z = 1$ for relics of $300 \times 100$~kpc$^2$ \citep{jones+23, Lee24}. Wide-area surveys will build the statistical samples needed to robustly calibrate the $P_\nu$--$M_{500}$ and $P_\nu$--LLS scaling relations, constrain the relic luminosity function, and map occurrence as a function of cluster mass, redshift, and dynamical state. Deep targeted observations will complement the surveys by providing the spectral and polarimetric detail needed to understand the physical drivers behind the observed correlations. Together, these programmes will establish whether the simulation overprediction reflects undetected faint relics or missing physics.

\subsection{Radio phoenices and revived fossil plasma}
\label{sec:phoenix}

Distinguishing between the competing revival mechanisms, pure adiabatic compression, gentle re-acceleration, or hadronic secondaries from relativistic protons producing $e^{\pm}$ pairs in magnetised filaments \citep{Keshet2025arXiv250307714Kphoenix}, requires broadband spectral data with sufficient frequency coverage and sensitivity to measure the curvature shape precisely, which current instruments cannot provide across the population. SKA-Low will be the primary tool for systematic identification and characterisation of radio phoenices. At 50--350~MHz, the spectral signature that distinguishes phoenices from relics, strong curvature from a re-energised aged population rather than a freshly injected power law, is most apparent, with the spectral turnover from adiabatic compression producing a characteristic concave spectrum \citep{Ensslin2001A&A...366...26Eaddicomp}. SKA-Low's continuous frequency coverage will sample this shape with sufficient resolution to constrain the compression factor and fossil plasma age. Complementary SKA-Mid observations will measure the steep high-frequency cutoff and enable spatially resolved spectral mapping to test whether pure compression suffices or additional re-acceleration is required. For the first time, SKAO surveys will provide an unbiased census of revived fossil plasma sources across a large cluster sample, constraining the volume-filling fraction of fossil plasma in the ICM and the duty cycle of AGN activity, quantities that feed directly into models of seed electron availability for radio relics and halos.

\section{Science verification with merger shocks}\label{sciverification}

Thanks to the extensive studies conducted with SKA precursors and pathfinders, several bright and well-characterized clusters have emerged as ideal science-verification targets for SKA observations of merger shocks and radio relics. These include El Gordo (ACT-CL J0102$-$4915, z = 0.87), Abell 3667 (z = 0.055), the Bullet Cluster (1E 0657$-$56, z = 0.296), PLCKG287.0+32.9 (z = 0.39), Abell 521 (z = 0.245). Together, they span a wide range in redshift, morphology, and dynamical state, providing excellent benchmarks to test the SKA’s imaging fidelity, sensitivity to diffuse low-surface-brightness emission, and polarimetric precision. El Gordo probes the high-z regime where inverse-Compton losses are severe; whereas Abell 3667 serves as the nearby archetype for calibrating wide-field imaging of bright double relics; the Bullet Cluster provides a textbook example of a strong bow shock with a well-measured Mach number; and Abell 521 represents a 2 Mpc relic with a distinct surface brightness distribution along its arc with a bridge emission between its radio halo and relics. Early SKA observations of these systems will be crucial for validating calibration strategies, testing theoretical models, and establishing the observational foundation for future large-scale studies of merger shocks and radio relic populations.
For the case of phoenix sources, the phoenix hosted in the cluster Abell 4038 is possibly the most promisng as it has been well characterized with the GMRT \citep{2018MNRAS.480.5352K} and MeerKAT \citep{kolokythas+25}.

Radio relics stand as exceptional laboratories for investigating particle acceleration and magnetic field amplification at cluster merger shocks. The wealth of observational data from SKA pathfinders has firmly established their connection to merger-driven shocks through morphological alignment, spectral signatures, and energetic arguments. There are other phenomena such as radio phoenices and gently re-energised tails which are related to discontinuities in the ICM which also offer different probes. Yet detailed microphysics, from seed particle populations to the mechanisms driving magnetic field growth, remains elusive.

The SKAO will decisively address these questions. Its sensitivity and resolution will permit resolved spectro-polarimetric studies of individual relics across cosmic time, from nearby archetypes to high-redshift systems, while simultaneously building large statistical samples to constrain population properties and their evolution. Wideband observations spanning 50 MHz to 1.76 GHz will map magnetic field organization through detailed RM synthesis and depolarization analysis, distinguishing shock-compressed ordering from turbulent amplification. High-resolution imaging will illuminate the filamentary structures that encode shock microphysics, testing whether acceleration is distributed or localized.

Science verification with well-studied clusters, particularly Abell 3667, Bullet, Abell 521, and El Gordo, will establish the observational foundation. For radio phoenices, Abell 4038 is a potential target. Subsequent targeted studies and wide-area surveys will probe the full diversity of shock-influenced sources and their cosmological context. Through this program, SKAO will transform our understanding of particle acceleration at the most violent events in the Universe, with implications extending from fundamental plasma physics to the energetics of large-scale structure formation.


\bibliographystyle{abbrvnat-maxbibnames4}
\bibliography{chapter,GDGbib,ap} 

\end{document}